\documentclass[
twocolumn,
]{ceurart}

\AtBeginDocument{%
  }
\usepackage{listings}
\usepackage{graphicx}
\usepackage{multirow}
\usepackage{verbatim}
\usepackage{float}
\usepackage{subcaption}
\usepackage{mathrsfs}
\begin{document}


\conference{The 1st Workshop on Risks, Opportunities, and Evaluation of Generative Models in Recommender Systems (ROEGEN@RECSYS'24)}

\title{A Prompting-Based Representation Learning Method for Recommendation with Large Language Models}

%\author[1]{Junyi Chen}[%
%orcid=0009-0009-4283-0663,
%email={chen-junyi666@g.ecc.u-tokyo.ac.jp,
%]
%\author[2]{Toyotaro Suzumura}[%
%orcid=0000-0001-6412-8386,
%email={suzumurat@gmail.com,
%]
\author[1]{Junyi Chen}[%
orcid=0009-0009-4283-0663,
email=chen-junyi666@g.ecc.u-tokyo.ac.jp,
]
\address[1]{The University of Tokyo, Tokyo, Japan}
\author[1]{Toyotaro Suzumura}[%
orcid=0000-0001-6412-8386,
email=suzumura@acm.org,
]
% \address[2]{The University of Tokyo, United States}
\cormark[1]
\cortext[1]{Corresponding author.}

\begin{abstract}
In recent years, Recommender Systems (RS) have witnessed a transformative shift with the advent of Large Language Models (LLMs) in the field of Natural Language Processing (NLP). Models such as GPT-3.5/4, Llama, have demonstrated unprecedented capabilities in understanding and generating human-like text. The extensive information pre-trained by these LLMs allows for the potential to capture a more profound semantic representation from different contextual information of users and items.

While the great potential lies behind the thriving of LLMs, the challenge of leveraging user-item preferences from contextual information and its alignment with the improvement of recommender systems needs to be addressed. Believing that a better understanding of the user or item itself can be the key factor in improving recommendation performance, we conduct research on generating informative profiles using state-of-the-art LLMs.

To boost the linguistic abilities of LLMs in recommender Systems, we introduce the \textbf{P}rompting-Based Representation Learning Method for \textbf{R}ecommendation (P4R). In our P4R framework, we utilize the LLM prompting strategy to create personalized item profiles. These profiles are then transformed into semantic representation spaces using a pre-trained BERT model for text embedding. Furthermore, we incorporate a Graph Convolution Network (GCN) for collaborative filtering representation. The P4R framework aligns these two embedding spaces in order to address the general recommendation tasks.
In our evaluation, we compare P4R with state-of-the-art recommender models and assess the quality of prompt-based profile generation.
\end{abstract}

\begin{keywords}
  Large language models \sep
  Prompting \sep
  Graph Convolutional Network \sep
  Recommendation
\end{keywords}

\maketitle

\section{Introduction}
Recommendation systems are pivotal in aiding users to discover pertinent and personalized items or content. Graph-based approaches have witnessed great success in recommender system development. The GCN-based methods such as NGCF\cite{ngcf}, LightGCN\cite{lightgcn},improve the generally accepted collaborative filtering recommendation baselines in a manner that it could learn user and item embeddings by linearly propagating them on the user-item interaction bipartite graph. Although the interaction matrix used to represent the interactions between users and items is data efficient and direct, it ignores rich textual information hidden behind the candidate raw data (eg. user reviews, professions, item locations, categories, ,etc) that could exert potential benefits to recommenders. Furthermore, which auxiliary information can be used for a recommender system is not researched thoroughly. Recently, the advent of Large Language Models (LLMs) in Natural Language Processing (NLP) has sparked increased enthusiasm for leveraging the capabilities of these models to elevate and improve recommendation systems.

With the thriving of pre-training in NLP, many language models have been pre-trained on large scale unsupervised corpora, and then fine-tuned for downstream tasks. The transformer architecture\cite{2017attention}, was introduced in 2017, has become a foundation in LLMs. 

%%It eschewed the sequential nature of Recurrent Neural Networks (RNNs) in favor of a self-attention mechanism, enabling parallelization and significantly improving efficiency in handling sequential data. 

Based on Transformer architecture, many Pre-trained Language Models (PLMs) have emerged. GPT series\cite{dale2021gpt}, developed by OpenAI, and BERT\cite{devlin2018bert}, developed by Google, represent two prominent approaches to leverage transformers for PLMs. The key advantage of incorporating PLMs into recommendation systems lies in their ability to extract high-quality representations of textual features and leverage the extensive external knowledge encoded within them.\cite{2023arXiv230203735L} One of the current focuses of research in the LLM-based recommendation systems is that how to align recommendation approaches with the characteristics of the ability of LLMs through specially designed prompts. Different from traditional recommendation systems, the LLM-based models can capture contextual information, comprehending user queries, item descriptions, and other textual data more efficiently.\cite{P5} Based on PLMs, fine-tuning strategy involves training the model on a smaller task-specific dataset. This dataset is typically related to a specific application or domain, such as sentiment analysis, text classification, question answering, or recommender systems. However, fine-tuning large language models on specific downstream tasks paradigm usually needs to fine-tunes all of the parameters in a PLM, which is a computational resource consuming process. Most researchers and companies cannot access as much resource as OpenAI or Microsoft or Google. As a result, a recently proposed paradigm, prompt learning\cite{liu2023pre}, further unifies the use of PLMs on different tasks in a simple yet flexible manner. 

In recent times, prompting for recommender systems has gained more attention. In general, prompting is the process of providing additional information for a trained model to condition while predicting output labels for a task, for example, providing a piece of text inserted in the input examples. Prompt tuning bridges the gap between pre-training and downstream objectives, allowing better utilization of the rich knowledge in pretrained models. This advantage will be multiplied when very little downstream data is available. Only a small set of parameters are needed to tune for prompt engineering, which is more efficient\cite{surveyllm4rec}. For example, Pretrain, Personalized Prompt, and Predict Paradigm (P5)\cite{P5} is a unified text-to-text paradigm for various recommendation tasks. It performs various tasks in an NLP manner using pre-trained prompting systems. However, a unified framework has limitations, such as not giving sufficient attention to personalized information feature representations using prompts. This ultimately hampers the performance of recommender systems. Another approach InstructRec\cite{instructrec} treats recommendation tasks as an instruction question answering task. These methods empower LLM's ability to understand the instructions in recommendations, however, such methods require complex pre-training or fine-tuning designs. When facing different recommendation situations, they usually deteriorate. Due to their lower efficiency and flexibility compared to existing recommendation systems, these methods need to be improved when adopting to the actual industry. 

\textbf{Contributions}
Our approach aims at a more affordable, yet efficient LLM enhanced recommender system. To achieve this, we propose a Prompting-Based Representation Learning Method for Recommendation  (P4R). The core idea is to utilize the inference ability of LLMs on limited item information, and utilize the GNN-based item-side collaborative filtering techniques. Our main contributions can be summarized as follows:
\begin{itemize}
\item The Connection. By connecting the GNN-based collaborative filtering recommendation framework with the open-sourced LLama-2-7b large language model, the P4R can improve its performance by enhancing the current item profile representations. This approach outperforms most GNN-based light-weight recommender systems and is particularly suitable for smaller companies or organizations. 
\item The Representation. The utilization of the BERT pre-trained model architecture connects the Natural Language Generation task with general recommendation tasks by adapting the embedding vectors of the generated profiles. Moreover, incorporating LLM-enhanced embeddings proves advantageous for representation learning. 
\item The Prompting. We have proposed a contextual attribute-aware prompting format to generate informative item profiles using LLMs. Our method emphasizes that intrinsic and extrinsic textual information should be treated differently when using prompting. This method demonstrates flexibility. Without extensive pre-training of large language models, it is capable of handling various candidate textual information.
\item The Performance. We compare P4R with different state-of-the-art recommender models and verify the performance of our approach. Furthermore, we examine which contextual information is crucial for LLM-based recommenders. We also conduct ablation study to analyze different designs for LLM-based recommender systems.
\end{itemize}
\section{RELATED WORK}
\subsection{GNN-based Collaborative Filtering}
Collaborative Filtering (CF)\cite{cf} has achieved great success in recommender systems since it was proposed. It has become the fundamental techniques in this field. In recent years, with the rise of Graph Neural Networks (GNNs), many GNN-based CF approaches become popular in this domain. For example, Neural Graph Collaborative Filtering (NGCF)\cite{ngcf}exploits the user-item graph structure by propagating embeddings on an interaction bipartite graph, and verifies the importance of embedding propagation for learning better user and item representations. LightGCN\cite{lightgcn} further demonstrates the effectiveness of Graph Convolutional Network structure in recommendation fields and include the most essential component in GCN: neighborhood aggregation for collaborative filtering. 
\subsection{Textual Representation learning enhanced Recommendation}
%%In the field of Natural Language Processing (NLP), a word can be represented as a vector in Euclidean space, and in Euclidean space, the vectors of synonymous words should also be closer in distance. Similarly, for graphs, a node can also be represented as a vector, and nodes with closer relationships should be closer in Euclidean space as well. The process of learning specific representations for such purposes is called representation learning.

Representations can entangle and hide more or less the different explanatory factors of variation behind the data\cite{representationlearning}, and contribute to the general improvement of machine learning based algorithms. Representation learning enhanced recommendation has gained much popularity with the advent of deep learning techniques, which allow for the extraction of complex and high-dimensional features from diverse types of data. The pure ID-indexing based recommending approaches suffer from discrete representation of user-item interaction, and ignore the rich side information contained in the contexts of users or items. As a result, some approaches explore to leverage these meaningful context representations for a better recommender construction. Unisec\cite{unisec}is one such approach that takes advantage of item descriptions to learn transferable representations from various recommendation scenarios\cite{surveyllm}. To be more specific, relative to the traditional modeling based on item IDs, the author primarily utilizes textual descriptions associated with products to model them, thereby mitigating the limitation of discrete representations.
\begin{figure*}[h]
  \centering
  \includegraphics[width=\linewidth]{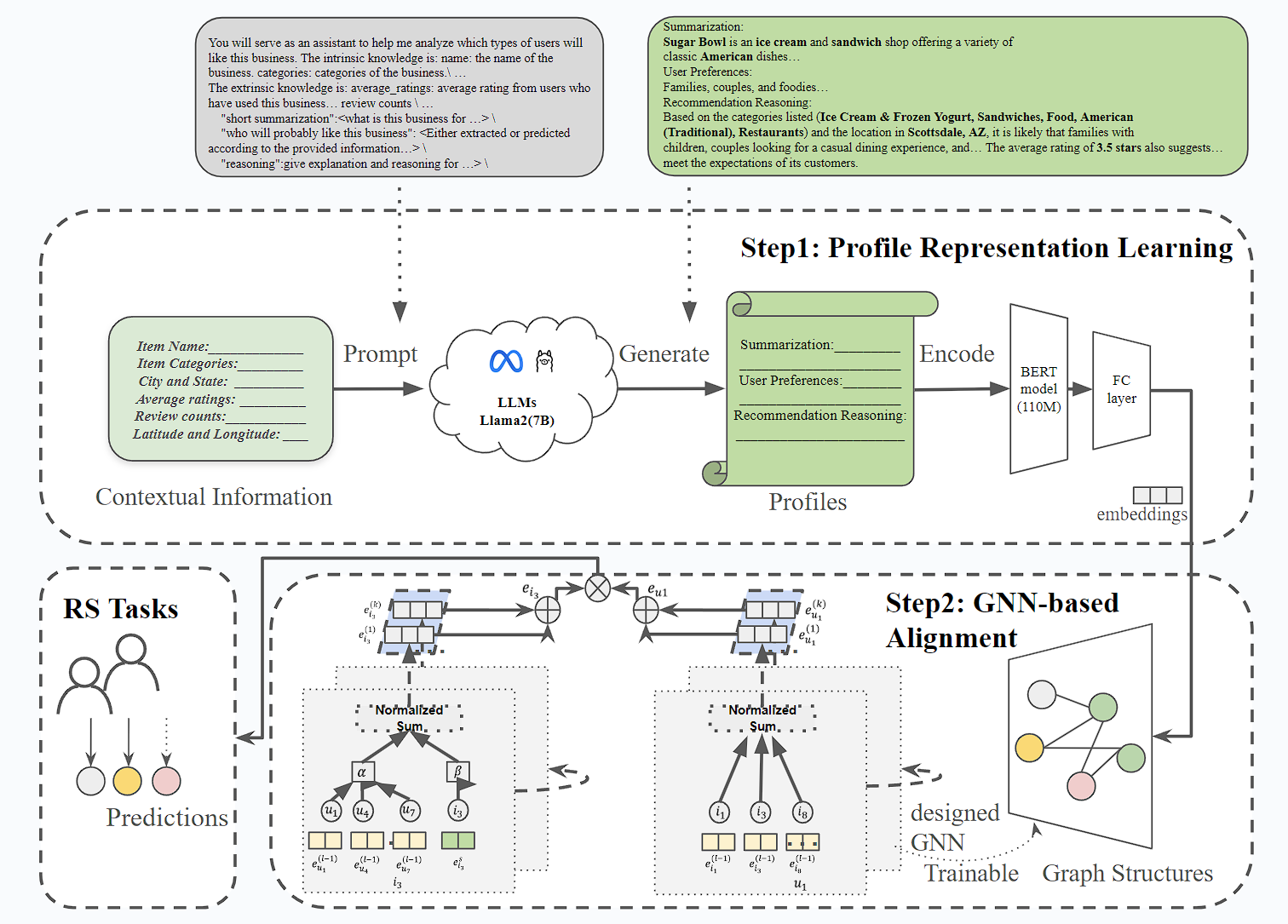}
  \caption{Our P4R framework learns a better item textual-based profile representation with the latest LLM prompting method. And it aligns with Graph Convolutional Network based collaborative filtering recommendation node representations in a unified training process.}

\end{figure*}
\subsection{Tranformer-based Recommendation}
Researchers propose Transformer-based recommendation approaches. For example, BERT4Rec\cite{sun2019bert4rec} models sequential user behavior with a bidirectional self-attention network through Cloze task. Transformer4Rec\cite{de2021transformers4rec} adopts a huggingface transformer-based architecture as the base model for next-item prediction. These two models have laid the foundation for Transformer-based recommender systems. 
\subsection{LLM-based Recommendation}
P5\cite{P5} is pre-trained on T5\cite{T5} model as its backbone. It formulates the main recommendation tasks such as sequential recommendation, rating prediction, explanation generation, review summarization, and direct recommendation tasks as a text-to-text problem. The model is pre-trained on multitask prompts for recommendation datasets in advance. Other researchers argue that fine-tuning or parameter efficient fine-tuning LLMs for downstream recommendation tasks is more efficient and saves the computational resources. For example, Petrov and Macdonald proposed GPTRec\cite{gptrec}, which is a generative sequential recommendation model based on GPT-2. In parameter fine-tuning training framework, only a small proportion of parameters are trained on some specific datasets, TallRec\cite{bao2023tallrec} introduces an effective framework on the Llama-7B model and LoRA(Low-Rank Adaptation) for aligning LLMs with recommendation tasks. However, all these methods apply LLMs to recommendation tasks in a manner that LLMs is directly used as a recommender itself. It is inevitable that these methods introduce additional noise and hallucinations into the system. Furthermore, the substantial training overhead and slow inference time restrict their practical application. 

%%To overcome this challenge, our approach utilize the prompting strategy to generate item semantic representations on an open-sourced lightweight Llama-2-7B model, we align the LLM-enhanced textual information representation side with the GNN-based collaborative filtering side for an effective general recommendation. 

\section{METHODOLOGY}
\subsection{Problem Setup}

We divide the recommendation problem into two main sub-tasks. One is to learn a better item textual-based profile representation with the latest LLM prompting method. Another is to align these representations with Graph Convolutional Network based collaborative filtering recommendation node representations in a unified training process. The overall framework is depicted in Figure 1. To be specific, in a recommendation scenario, we have the user set $\rm{U}=\{u_1,u_2,...,u_n\}$, the item set $\rm{I}=\{i_1,i_2,...,i_m\}$ and the rating score set $\rm{R}=\{r_{11},r_{12},...,r_{nm}\}$, where $r_{ij}$ is collected from the users' past rating behaviors. When no interaction between a user-item pair $(u_i,i_j)$ exists, the value is set as 0, otherwise it should be integers from 1 to 5.
Furthermore, the auxiliary textual information set which contains the descriptions of the items is expressed as $\rm{T}$. The goal is to learn a LLM-enhanced user and item representations $e_u$ and $e_i$ through the model. 
\subsection{Proposed Method}
\subsubsection{Auxiliary Feature Extraction through In-context learning}
Believing that the essence hidden behind product descriptions is not fully incorporated, our approach introduces textual information such as item names, categories, and locations. This integration enhances the understanding of user preferences. Alongside the introduction of GPT-3\cite{gpt3}, In-Context Learning (ICL) is proposed as an advanced prompting strategy, which significantly boosts the performance of LLMs on adapting to many downstream tasks\cite{surveyllm}. The difference between this method and traditional machine learning strategy is that it does not optimize any parameters. It demonstrates a performance comparable to fine-tuning or pre-training for downstream tasks, while also exhibiting advantages in terms of resource comsumption and efficiency.

Enlightened by the Google research teams' Chain-of-thought (CoT)\cite{COT} reasoning prompt design, we propose a recommendation-oriented reasoning prompting format $\mathscr{P}$ that explores the potential of LLMs reasoning abilities by generating item profiles $\mathbb{I}$. The outline of the process is as follows:

\begin{equation}
\mathbb{I}=\mathscr{P}(\rm{I_j}, \mathit{T_j})    
\end{equation}

A series of specially designed reasoning steps are provided to assist LLMs analyze the nature of the input candidates. For item prompt construction, we categorize the textual information $\mathit{T_j}\in \rm{T}$ for an item $\mathit{i}\in \rm{I^{m}}$
into \textbf{Intrinsic Information Attribute $\mathit{t_i}$} and \textbf{Extrinsic Information Attribute $\mathit{t_e}$}: 
\begin{equation}
    \rm{T_j}=\{ \mathit{t_i},\mathit{t_e}\}
\end{equation}
The intrinsic information attribute $\mathit{t_i}$ includes the direct and accurate textual information collected from the candidate itself, such as item title/name, categories, brand, locations, etc. This information is usually clearly shown and unlikely to change over a long range of time, which lays the base foundation of an item's nature. And we argue that intrinsic attributes should be paid more attention in generating a whole-picture of item profiles, for they are the direct expression of the item properties which contain less biases or misleading information. The extrinsic information attribute $\mathit{t_e}$ includes the textual information that does not connect to the item directly, usually the subjective feedback (eg, average ratings, selected reviews, review counts, etc). Although personal will unavoidably influences the extrinsic attributes, it is important to include feedback when creating item profiles using prompts that consider diversity. By designing prompts appropriately, language models can mitigate the biased information caused by personal will and extract valuable information to enhance the generation of item profiles. Our proposed prompt format include three core concepts, they are: \textbf{The General Summary} of the item attributes, \textbf{The User Preference Predictions} leveraged by LLMs rich background knowledge, and \textbf{The Recommendation-Oriented Reasoning} details for supporting aforementioned two core parts. The summarization part is a common NLP task and is the advantage of LLMs. It can capture the intrinsic knowledge of items, so we consider this part to be the most essential in our prompt format. The preference prediction part is an application that can make use of the extensive amount of pre-trained texts in PLMs. This approach differs from traditional machine learning methods because PLMs use very large corpora that have hidden benefits not attainable with limited datasets used in traditional methods. The final part of the reasoning process plays a crucial role in providing a compelling explanation for the recommendation. We utilize the recommendation-oriented reasoning prompting strategy to guide the LLMs in understanding the significance of specific inputs. For example, provide brief explanations for several criteria. By instructing the LLMs to analyze this information and construct a self-contained logical reasoning process, we enable researchers to gain insights into the underlying logic behind the generated results. Furthermore, this component aids in selecting the most suitable techniques for feature extraction engineering and representation learning.

%%\begin{figure}[h]
%%  \centering
 %% \includegraphics[width=\linewidth]{figure1.png}
%%  \caption{The overall depiction for item profile generation}
%%  \Description{Item Profile Generation}
%%\end{figure}

\subsubsection{Textual embedding and representation}
Our goal is to obtain an informative representation that benefits Item-based CF recommender models. Through the generated item profiles, we have leveraged the important item features from massive contextual information. It is crucial to align this highly-correlated knowledge with CF-based recommendation item representations.  
As Bidirectional Encoder Representations from Transformers (BERT)\cite{devlin2018bert} has achieved great success in utilizing Transformer architecture in NLP field\cite{2017attention}, Masked Language Modeling (MLM) is an essential technique to predict words by randomly masking a certain percentage of the corpus in the input side, then run the entire masked sentence through the model. We import a similar Pre-trained BERT model as our text-embedding model. The generated profile texts are fed to the pre-trained BERT model to learn the semantic representation of each significant item profiles. For each item profile, each part of the textual information is encoded, following the BERT initial design. In Figure 2, the initial semantic embeddings of each item are obtained. $e_{init}$ for each profile is obtained through BERT encoders, and through the PLM, the hidden states are obtained from the last layer $\mathit{L_{12}}$ of the BERT model. For embedding dimension customization, then the CLS token is connected to a Fully Connected Layer $\mathit{FCL}$ and a Rectified Linear Unit layer $\mathit{ReLU}$ to get the tailored sized representations. This process is necessary because directly using the embeddings obtained from the BERT layers often leads to redundancy, which can affect the subsequent operations.

\begin{figure}[h]
  \centering
  \includegraphics[width=0.25\textwidth]{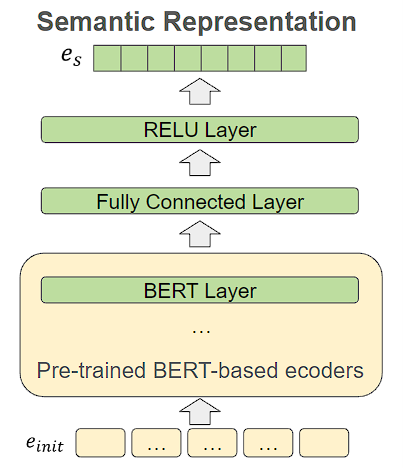}
  \caption{A BERT-based item semantic representation extraction framework}
  
\end{figure}

\subsubsection{Alignment with recommendation through GNN-based approach}
GNN-based Item Collaborative Filtering has shown its advantages in leveraging the relationship between items and users. In a traditional GCN model, user and item embeddings are learned by linearly propagating them through the user-item interaction graph. We inherit this GNN-based learning mechanism. Let $e_u$ and $e_i$ denote the embeddings of users and items separately, $e_s$ denotes the LLM-enhanced item embedding, we adopt the crucial GCN components -- neighborhood aggregation\cite{lightgcn} as follows:
\begin{equation}
    e_u^{(k+1)}=AggregationFunc(e_u^{(k)},e_i^{(k)} \in N_u)
\end{equation}

\begin{equation}
    e_i^{(k+1)}=AggregationFunc(e_i^{(k)},e_u^{(k)} \in N_i)
\end{equation}
$AggregationFunc$ is the aggregation function which considers the k-th layer's representation of the target node and its neighbor nodes. The $N_u$ denotes the set of items that are interacted by user $u$, $N_i$ denotes the set of users who have interacted with item $i$. The operation of learning user and item representation for recommendation is as follows: 

\begin{equation}
    e_{i}^{(k+1)}=\sum_{u\in N_i} \frac{1}{\sqrt{N_u} \sqrt{N_i}} [\alpha( e_{u}^{(k)}+ \beta e_s)]
\end{equation}
\begin{equation}
    e_{u}^{(k+1)}=\sum_{i\in N_u} \frac{1}{\sqrt{N_u} \sqrt{N_i}} e_{i}^{(k)}
\end{equation}

where $e_{u}^{(k)}$ and $e_{i}^{(k)}$ respectively denote the embedding of user and item after k-th layer propagation. The $\alpha$ and $\beta$ are controllable hyper-parameters. 
$\frac{1}{\sqrt{N_u} \sqrt{N_i}}$ is the GCN symmetric normalization\cite{GCN}. The final representation for user and item is obtained by calculating the sum of embeddings from each layer of the models. 
%%\begin{figure}[h]
%%  \centering
%%  \includegraphics[width=\linewidth]{figure4.png}
%%  \caption{The illustration of GCN model architecture. $e_s$ denotes the LLM-enhanced item embedding generated from system prompt approach. We sum all the embeddings from each layers to get the final presentation of each node.}
%%  \Description{}
%%\end{figure}
We employ the Bayesian Personalized Ranking (BPR) loss\cite{rendle2012bpr} to train the embeddings of the layer. BPR is a pairwise loss that is used in recommendation systems to optimize models for personalized ranking, aiming to improve the ranking of relevant items for each user. For a user historical behavior triple $(u,i,r)$, where $u$,$i$,$r$ denote user, item, and the rating respectively, the goal of the BPR loss is to learn an item ranking function for each user that sorts items according to the user's preferences:
\begin{equation}
    \mathcal{L_{BPR}}=-\sum_{(u,i,j)\in D}ln\sigma (f_u(i)-f_u(j))
\end{equation}
where $D$ denotes all triples from the history, $f_u(i)$ denotes the preference score of user $u$ to item $i$. Item $j$ is randomly picked from the user's unseen items in the item set, and considered as a negative item. The BPR loss function can be minimized by optimization methods such as SGD\cite{sgd}, Adam\cite{kingma2014adam} to learn the appropriate item ordering function $f_u(i)$.
\section{Experiments and Results}
In this section we evaluate our proposed method on open datasets Yelp$ \footnote{https://www.yelp.com/dataset}$ and Amazon Video Games$\footnote{https://cseweb.ucsd.edu/~jmcauley/datasets}$, and present a series of ablations.
\subsection{Experimental Setup}
\subsubsection{Datasets}

We perform the evaluation on subsets collected from Yelp2018 and Amazon-VideoGames datasets. Both datasets are popular in recommendation tasks, there are over 6 million interactions from only Yelp dataset in total and over 2.5 million interactions from Amazon-Video-Games dataset, it is important to collect a sub-dataset for the affordable research. The candidates are randomly selected based on their average interaction times. Specifically, users from the Yelp2018 subset have an average of over forty interactions per user, while each item is interacted with an average of five times by users. A similar approach is applied to the Amazon dataset. The overall sparsity of the subsets are both collected around 99\%, which is the common case for recommendation data collections. The statistics are presented in Table 1. Our data and implementations are available in Anonymous github\footnote{https://anonymous.4open.science/r/P4R}.

\begin{table*}[ht]
  \caption{Statistics of the experimented data}
  \label{tab:freq}
  \begin{tabular}{c|c|c|c|c}
    \toprule
    Dataset&User &Items &Interaction &Sparsity\\
    \midrule
    Yelp2018 &767 & 3647&27453 &99.018571\%\\
    Amazon-VideoGames&795&6627&37341&99.291235\%\\
  \bottomrule
\end{tabular}
\end{table*}

\subsubsection{Implementation Details}
The framework is a two-step process, the first step is profile representation learning, and the other step is the GNN-based training process. It is more flexible than an end-to-end training model.

We first implement the prompt-based profile generations with Llama-2-7b-chat model\cite{touvron2023llama} on HuggingFace library, it is a fine-tuned version of the Llama-2-7b model with over 7 billion parameters. Llama 2 is an auto-regressive language model that uses an optimized transformer architecture. The tuned versions use supervised fine-tuning (SFT) and reinforcement learning with human feedback (RLHF) to align to human preferences. For generation, we simply use the greedy decoding method. The number of highest probability vocabulary tokens to keep for top-k-filtering is set to 30. Max generated tokens length is 512. Then, for textual embedding and representation of the synthetic profiles, we employ the pre-trained BERT model-110M on HuggingFace library which have 12 layers in total. For each sentence, 15\% of the tokens are masked. The masked tokens will be replaced randomly on a certain percentage. And we connect the final hidden states' CLS token to a Fully connected layer and Relu layer to get the customized embeddings. We set the generated embedding dimensions to 32, 64, and 128, and evaluate them separately. During the training operation, we use Adam learner to minimize the training loss of BPR function. The default layer number is two, and the default learning rate is $1\times 10^{-3}$, training batch size is 2048. We also employ the early stop techniques for training. The ordering strategy is random ordering. The training and evaluation is conducted using one RTX 3090 series GPU, and we implement it on RecBole\footnote{https://recbole.io/} platform.

\subsubsection{Baselines}
We compare P4R model with several well accepted baselines for general recommendation tasks. \underline{NGCF}\cite{ngcf}: Neural Graph Collaborative Filtering is the most well accepted baseline in GCN-based recommendation system. It lies the foundation of the GNN-based CF method. And it performs better than tradional CF method. \underline{LightGCN}\cite{lightgcn}: A Graph Convolutional Network (GCN) baseline for recommendation systems. It is a light-weight variant of traditional GCNs that aims to address the scalability and performance issues faced by GCNs in recommendation tasks. \underline{SGL}\cite{SGL}:A Self-supervised Graph Learning for Recommendation , designed to supplement the classical supervised task of recommendation with an auxiliary self-supervised task, which reinforces node representation learning via self-discrimination. 

\subsection{Results}
\subsubsection{Overall Performance}

Table 2 presents the performance of the proposed model compared
against baselines. 

\begin{table*}[ht]
  \begin{center}
    \caption{The perfomance of the baselines and proposed model P4R}
    \begin{tabular}{|c|c|c|c|c|c|c|c|c|c|c|c} 
      \hline
      \textbf{Dataset} & \multicolumn{8}{c|}{\textbf{Yelp}}\\
      \hline
      \textbf{Model}& Recall@10& Recall@20&NDCG@10&NDCG@20&MRR@10&MRR@20&Hit@10&Hit@20\\
      NGCF & 0.1060 & 0.1565&0.0861&0.1034&0.1374&0.1447&0.2641&0.3686\\
      LightGCN & 0.1796&0.2266&0.1615&0.1777&0.2512&0.2572&0.4000&0.4889\\
      SGL & 0.1872 &0.2536 &0.1725&0.1940&0.2750&0.2827&0.4275&\underline{0.5386}\\
      P4R &\underline{0.2030}&\underline{0.2599}&\underline{0.1823}&\underline{0.2009}&\underline{0.2813}&\underline{0.2872}&\underline{0.4444}&0.5307\\ 
      \hline
            \textbf{Dataset} & \multicolumn{8}{c|}{\textbf{Amazon-VideoGames}}\\
      \hline
      \textbf{Model}& Recall@10& Recall@20&NDCG@10&NDCG@20&MRR@10&MRR@20&Hit@10&Hit@20\\
      NGCF & 0.0639 & 0.0880&0.0676&0.0754&0.1391&0.1438&0.2322&0.3020\\
      LightGCN & 0.1131&0.1449&0.1148&0.1245&0.2153&0.2203&0.3477&0.4213\\
      SGL & 0.1340 &0.1653 &0.1347&0.1442&0.2438&0.2479&0.4023&0.4569\\
      P4R &\underline{0.1411}&\underline{0.1770}&\underline{0.1569}&\underline{0.1674}&\underline{0.2954}&\underline{0.2988}&\underline{0.4048}&\underline{0.4695}\\ 
      \hline
    \end{tabular}
  \end{center}
\end{table*}

 The evaluation metrics are $Recall$,$NDCG$,$MRR$,and $Hit$, they denote the Recall score, Normalized Discounted Cumulative Gain (NDCG), Mean Reciprocal Rank (MRR), and Hit rate respectively. This table shows the perfomance of different baselines and our proposed method when the embedding size is all set as 64. P4R sees statistically improvement compared to other baselines. In the best validation test result, it outperforms the other best baselines by 8.4\% and 7.0\% on $Recall@10$ and $Recall@20$ measures respectively, 16.5\% and 16.1\% on $NDCG@10$ and $NDCG@20$ measures, 21.2\% and 21.0\% on $MRR@10$ and $MRR@20$ measures, 4.0\% and 2.8\% on Hit rate measures. This demonstrates the value of the proposed method.
 \subsubsection{Profile Correlation Evaluation}
 We assume a highly correlated item profile generally benefits the understanding of user preferences and the final recommendation results, to validate our generated profiles align well to the original textual information, we employ Recall-Oriented Understudy for Gisting Evaluation (ROUGE) score\cite{lin2004rouge}. We report the result in Table 3. The high $ROUGE-1_{Recall}$ score shows the correlation between the raw data and the synthetic data and proves the generated contexts are informative. The original descriptions for items are generally short and vary from dataset to dataset, which will result in different ROUGE-scores. It should be noted that ROUGE scores are not the sole criteria for evaluating these generated profiles. We also aim for the prompt-based generated profiles to not only summarize the original content but also to utilize the reasoning and prediction capabilities of LLMs. Our generation task differs from traditional translation or explanation generation tasks LiPEPLER\cite{LiPEPLER}. Therefore, the BLEU score is not suitable for evaluating the result. We will leave the further evaluation to future researchers. 
 
 %%We also attempted to create user profiles based on their previous reviews in order to predict their personality and improve recommendations. Our expectation was that by analyzing their writing styles and interactions with items, we could extract or predict their preferences. However, we found that the review texts were heavily biased in terms of writing styles. This means that different individuals have different thresholds for items, especially when it comes to reviewing things like restaurant food, etc. For example, some people may find hamburgers delicious while others may dislike fast food due to its unhealthiness. When generating one's profiles using a limited subset of reviews, it becomes challenging to capture all of his preferences. For instance, one user may frequently visit a nearby barber shop and also use services from a nearby McDonald's chain store. These two types of businesses show different preferences for one's lifestyles, but the training set might only include former information, while the latter is put into the validation or testing set. This can lead to inevitable biases in the end. Our experiment confirmed this phenomenon. As a result, we exclude user profiles for recommendations in our model, as the generated user profiles did not contribute to the improvement of the recommendation task. However, we believe that user profiles could still be valuable for understanding user preferences.
\begin{table}
  \caption{ROUGE-1 score of the generated profiles}
  \label{tab:freq}
  \begin{tabular}{ccc}
  \hline
  \textbf{Dataset}& \textbf{Yelp}&\textbf{Amazon-VideoGames}\\
  \hline
  $ROUGE-1_{F1}$&0.0017828&0.0006645\\
  $ROUGE-1_{Precision}$&0.0008928&0.0003324\\
  $ROUGE-1_{Recall}$&0.5925926&0.900000\\
  \hline
\end{tabular}
\end{table}
\subsubsection{Ablations}

\begin{table*}[ht]
  \begin{center}
    \caption{The ablation study of different P4R-based design}
    \begin{tabular}{|c|c|c|c|c|c|c|c|c|c|c|c} 
      \hline
      \textbf{Dataset} & \multicolumn{8}{c|}{\textbf{Yelp}}\\
      \hline
      \textbf{Model}& Recall@10& Recall@20&NDCG@10&NDCG@20&MRR@10&MRR@20&Hit@10&Hit@20\\
      P4R &0.2030&0.2599&0.1823&0.2009&0.2813&0.2872&0.4444&0.5307\\ 
      P4R-B32&0.1957&\underline{0.2636}&0.1624&0.1854&0.2434&0.2504&0.4379&0.5373\\
      P4R-B128&\underline{0.2055}&0.2477&0.2059&0.2204&\underline{0.3353}&0.3403&0.4536&0.5242\\
      P4R-B256&0.2045&0.2579&\underline{0.2068}&\underline{0.2244}&0.3350&\underline{0.3406}&\underline{0.4575}&\underline{0.5399}\\

      P4R-WIP&0.129&0.1805&0.0981&0.1164&0.1412&0.1482&0.2954&0.3961\\
      P4R-WT&0.0245&0.0322&0.019&0.0215&0.0258&0.0272&0.0522&0.0731\\
      Random&0.0016&0.0026&0.0010&0.0014&0.0017&0.0020&0.0089&0.0140\\
      \hline
      \textbf{Dataset} & \multicolumn{8}{c|}{\textbf{Amazon-VideoGames}}\\
      \hline
      \textbf{Model}& Recall@10& Recall@20&NDCG@10&NDCG@20&MRR@10&MRR@20&Hit@10&Hit@20\\
      P4R &0.1411&0.1770&0.1569&0.1674&0.2954&0.2988&0.4048&0.4695\\ 
      P4R-B32&0.1261&0.1653&0.1144&0.1281&0.2008&0.2064&0.3680&0.4505\\
      P4R-B128&\underline{0.1528}&\underline{0.1894}&0.1717&0.1827&0.322&0.3268&\underline{0.4302}&\underline{0.5000}\\
      P4R-B256&0.1492&0.1826&\underline{0.1731}&\underline{0.1831}&\underline{0.3385}&\underline{0.3427}&0.4226&0.4886\\

      P4R-WIP&0.0787&0.1023&0.0715&0.0801&0.1248&0.1289&0.236&0.2944\\
      P4R-WT&0.0219&0.0251&0.0169&0.0179&0.0219&0.0226&0.0480&0.0594\\
      Random&0.0027&0.0056&0.0012&0.0023&0.0012&0.0019&0.0052&0.0157\\
      \hline
    \end{tabular}
  \end{center}
\end{table*}
%%\begin{figure}[htbp]
%%\centering
%%\includegraphics[width=\linewidth]{figure5.png}
%%\caption{The comparison between different model designs in terms of recall, MRR, NDCG, and Hit metrics} 
%%\Description{}
%%\end{figure}

We present various ablated designs in Table 4, we tested the recommendation results without the LLM-enhanced Item Profiles (P4R-WIP), the different embedding choices for the model, eg, P4R-B32 (Bert embedding size=32),P4R-B128,etc, the P4R model Without Training a unified embedding for items (P4R-WT), and the random recommendation.

\underline{P4R-WT}: We believe that creating item profiles that are highly correlated, accurate, and predictive would be advantageous for recommender systems. Therefore, we conducted a test to directly predict candidates by using item profile initial embeddings. Based on the results in Table 4, we can observe that even without any training specifically for the recommendation task, P4R-WT is capable of inferring candidates for users and outperforms random recommendations. This demonstrates the effectiveness of prompt-based generated item profiles.

\underline{P4R-WIP}: Our proposed model is built upon the GCN method, making it logical to evaluate its recommendation capability. Even without the Item Profiles, P4R-WIP demonstrates the effectiveness of the GCN-based collaborative filtering approach. This finding supports previous research that has already confirmed the benefits of using this method to extract relationships between nodes.

\underline{P4R-B256, P4R-B128 and P4R-B32}: P4R-256,P4R-B128 and P4R-B32 have been tested. The number 32, 128 or 256 refers to the embedding size of PLM Bert's token embedding, as well as the user and item embedding size of the GNN model. For P4R-128, the embedding size is 128. According to Table 4, increasing the size of the embedding enhances the performance of our proposed model over time. We believe this is because, within a certain range, a larger embedding size is associated with a better understanding of the semantic information representation generated by LLMs. For example, as for NDCG and MRR, both two metrics rank a higher score when increasing the embedding size, and it shows consistency for different datasets. The P4R-B128 model shows best Recall score performance in two datasts. For Hit rate, both the embedding size of 128 and 256 show a better performance than lower embedding size. The maximum increases, respectively, are 27.3\% in NDCG score on the Yelp dataset, 51.5\% in NDCG score on the Amazon-VideoGames dataset, 38.0\% in MRR score on the Yelp dataset, and 68.6\% in MRR score on the Amazon-VideoGames dataset, when the embedding size is changed from 32 to 256. The maximum increases, respectively, are 5.0\% in recall score on the Yelp dataset and 21.2\% in recall score on the Amazon-VideoGames dataset. In terms of Hit score, the maximum increases are 4.3\% on the Yelp dataset and 16.9\% on the Amazon-VideoGames dataset.
The reason why the approach achieves a better performance on Amazon-VideoGames dataset probably lies in the item category influence. The Yelp dataset is a cross-domain business dataset, which means it contains different businesses with more than one category. While on the other hand, the Amazon-VideoGames dataset only contains video games. For instance, one user may frequently visit a nearby barber shop and also use services from a nearby McDonald's chain store. These two types of businesses show different preferences for one's lifestyles, but they all exist in Yelp dataset, making it harder to capture his preferences. However, we can still observe improvements in both datasets. It demonstrates that our approach can capture user preferences even when the categories themselves vary a lot.

\underline{Random}: The Random method removes all designs and simply makes random recommendations,  The result indicates that the dataset we chose is suitable for experimentation.

\subsubsection{Case Study}
\begin{figure*}[htbp]
  \centering
  \includegraphics[width=\linewidth]{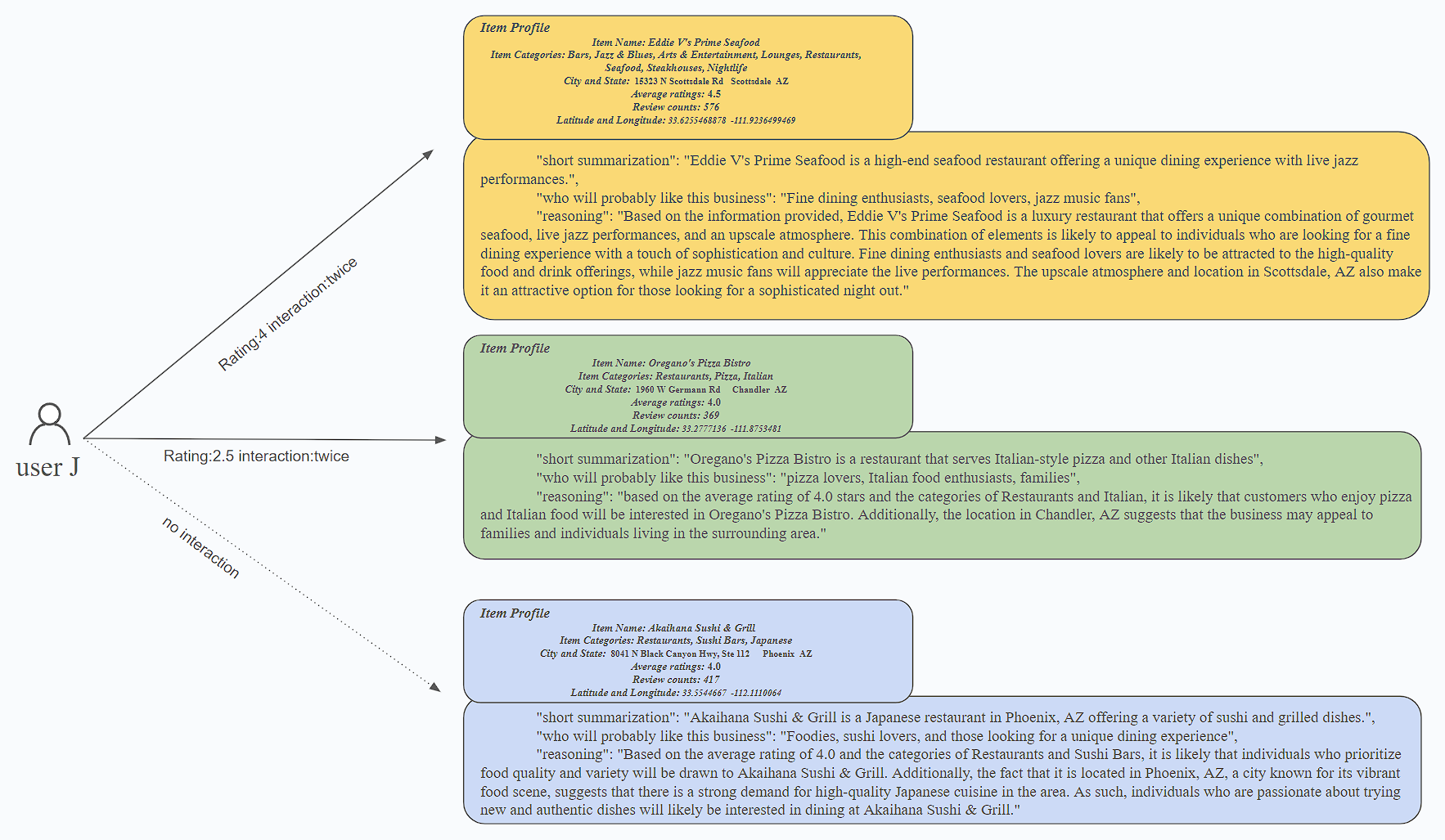}
  \caption{A Case Study for user preferred interaction. The user has different interactions with items. Each block shows a generated item profiles, it contains the item descriptions, which aligns with its intrinsic knowledge. The summarization, prediction, and reasoning for one item is also included in the profile.}
\end{figure*}
As shown in Figure 3, our method achieves a good performance in capturing the characteristics of different items and provides informative insights of user preferences. 

This example presents three profiles of different items from Yelp dataset, respectively. Each block shows the generated item profiles. Each items is close to each other, and they are from similar categories. It can be observed
that our approach provide additional item knowledge which is useful for predicting the user preferences. Specifically, for user J, his user history information shows he has been writing reviews on Yelp since 2012. He has written a total of 556 reviews, and the average rating is 2.93. The total interaction times for this user are 35, most of them are restaurant feedback. These three restaurants are arranged in descending order according to the predicted score of the user's preferences. The default rating for items without interaction is 0. Eddie V's Prime Seafood has the highest predicted score among these items.  According to his review, it mentions the restaurant's atmosphere, decorations, and the food quality, which is the case that the restaurant is categorized as Bars, Jazz and Blues, Arts and Entertainment, Lounges. From his past written reviews, we can tell that he is a picky customer (with an average rating of 2.93) who looks for fine dining experiences. He prefers American and Italian cuisine over Japanese cuisine. We can observe that this is consistent with the user's rating scores of the restaurant, indicating that our generated profiles can reflect the preference of potential users.
\section{CONCLUSION}
In this paper, we introduce P4R, a prompting-based approach to representation learning for general recommendation tasks. Our method leverages LLMs to extract features from item descriptions, generating informative and predictive item profiles. By incorporating a Graph Convolution Network based Collaborative Filtering measure, we demonstrate the effectiveness of aligning two domains for recommendations. We evaluate our models on datasets from Yelp and Amazon-VideoGames, and show that they outperform several strong baselines. Furthermore, through ablated models, we explore opportunities for further model improvement.

However, the P4R model does have some limitations. It primarily relies on historical interactions for generating content and does not consider sequential information when generating user profiles. As a result, it may not accurately capture long-term interests. %%In our proposed method, we chose not to use user profiles because the process of generating them introduces more noise and negatively impacts the results. One of the reasons for this is the inherent variability in user data, such as differences in attitudes, personalities, and reviewing styles. 

%%The current LLMs-based generation approach has not yet found a way to effectively address user biased preference differences. However, we believe that this is a valuable area for future research, and with improved user profiles, the overall performance of the model can be further enhanced.

There are various LLMs that need to be explored for prompting engineering. We utilized a Llama-2-7b model for our generation task because it is cost-effective and lightweight. However, it does have certain limitations when it comes to reasoning. It is believed that a larger model can generate more detailed profiles, but there is a challenge in finding the right balance between efficiency and accuracy. 

\section{Acknowledgments}

This work is supported by "Joint Usage/Research Center for Interdisciplinary Large-scale Information Infrastructures (JHPCN)" and "High Performance Computing Infrastructure (HPCI)" in Japan (Project ID: jh241004). In addition, this work was supported by JSPS KAKENHI Grant Number JP23H03408.

\bibliography{references}
\end{document}